\documentstyle[epsf,twocolumn,floats,prb,aps]{revtex}
\newcommand{\figwidth}{3.375in}
\begin{document}
\draft

\twocolumn[\hsize\textwidth\columnwidth\hsize\csname @twocolumnfalse\endcsname

\title{The Role of Substrate Corrugation in Helium Monolayer Solidification}
\author{M. E. Pierce and E. Manousakis}
\address{
Department of Physics and Center for Materials Research and Technology,
Florida State University, Tallahassee, FL 32306-4350
}
\date{\today}
\maketitle
\begin{abstract}
\noindent
We investigate the first layer of helium adsorbed on graphite with
path-integral Monte Carlo, examining the role of substrate corrugations
on the phase diagram.  When no corrugations are present, the equilibrium
state of the system is a liquid phase, with solidification occurring 
only under compression but before layer promotion.  We determine
the solid-liquid
coexistence region and compare our results to recent
Green's function Monte Carlo calculations on the same system.
When substrate corrugations are
included, we find that the equilibrium phase is the $\sqrt 3 \times \sqrt3$ 
commensurate solid phase that is well known from experiment.  The
melting behavior, heat capacity, and single particle binding energy
are determined and compared to experiment.  We further find that
for densities below the commensurate coverage, the low temperature phase
of the system consists of 
solid clusters in coexistence with coalesced vacancies.  We find  no
first layer liquid phase and so no superfluidity in this layer, in contrast to some
rather recent suggestions.  
\end{abstract}
] 

\section{Introduction} 
Quantum films such as helium adsorbed on graphite surface 
are characterized by a rich phase diagram as one varies the number  
of the deposited helium atoms.  As a function of the helium coverage, 
the helium film grows in atomically  thin  
layers\cite{dash75,schick80,bruch97}  and at least seven such   
layers may  be clearly distinguished and studied  on a well-prepared  
substrate.   
 
During the  last several years, extensive heat capacity 
measurements\cite{bretz77,greywall91,greywall93}  of   the  first  six 
layers have been performed, and superfluidity in the higher layers has 
been detected  by both torsional  oscillator\cite{reppy94,reppy96} and 
third sound measurements.    
More specifically, the first layer  of helium adsorbed on graphite has 
been  the laboratory for  study of  a variety  of phenomena.   In this 
layer  a $\sqrt  3 \times  \sqrt3$ commensurate  solid phase  forms in 
which  one-third  of  the  available substrate  adsorption  sites  are 
occupied. 
\cite{bretz77,greywall93,nielsen80,abraham87} 
For  densities  above  this  commensurate density,  the  monolayer  is 
characterized by  a region  of domain wall  phases and at  even higher 
monolayer  densities  it  forms  an  incommensurate  triangular  solid 
phase.\cite{carneiro76,hering76,rabedeau89}  The  phase diagram  below 
the commensurate  density at low  temperatures is not  well understood 
and there  are two competing  scenarios.  In one scenario  this region 
corresponds  to  a  solid  with  clustered  vacancies.   \cite{ecke85} 
According to  this picture, since  the commensurate solid phase  is in 
the same universality class as the three-state Potts model, 
\cite{bretz73,berker78,schick77,tejwani80,crary87} at lower densities 
the film should consist of  a commensurate solid with vacancies.  When 
the temperature is raised, the solid melts continuously.  Lowering the 
temperature causes  the vacancies to  coalesce, which is a  first order 
transition.   The difference  between  the temperatures  of these  two 
transitions becomes smaller as the  density is lowered until they meet 
at a tricritical point. 
 
The second scenario for the low density region 
of the phase diagram of the sub-monolayer was suggested more recently 
by Greywall and Busch \cite{greywall91} (GB). GB point out that their 
measured heat capacity is not linear 
in density for the entire region below the commensurate density, as it 
must be for phase coexistence.  Thus, they propose that a self-bound 
liquid phase  occurs at about  0.04 atom/$\AA^2$.  This  conclusion is 
supported  by  2D  variational calculations.  \cite{bruch93}  However, 
direct   measurements   \cite{reppy94,reppy96,mohandas95}  detect   no 
superfluidity, possibly because of poor substrate connectivity. 
 
In a  previous publication\cite{pierce99b}, we  have used the path-integral 
Monte  Carlo method  and realistic  helium-helium and  helium graphite 
interactions  to study  the  monolayer at  or  below the  commensurate 
density. We found that the  presence of corrugations, which causes the 
commensurate  solid  at   1/3  coverage,  creates  solid  commensurate 
clusters at densities below the commensurate density. We found that it 
is  the melting  of these  monolayer clusters  which gives  rise  to a 
specific heat  maximum which was  incorrectly 
interpreted\cite{greywall91} as  onset of 
monolayer superfluidity. In  this paper we examine in  detail the role 
of substrate  corrugation on the  first layer phase diagram  using the 
path-integral  Monte  Carlo method.   First,  we  present results  for 
helium   adsorbed  on   a  smooth   substrate.    The  helium-graphite 
interaction is based on the laterally averaged potential of Carlos and 
Cole.\cite{cole80} This layer exhibits liquid and solid phases, and we 
calculate  the phase diagram  at low  temperatures. These  results are 
compared     with    recent     Green's    function     Monte    Carlo 
calculations.\cite{whitlock98} The  second layer promotion  density is 
also determined.  Next,  calculations including substrate corrugations 
are  discussed,  and  direct  comparisons  of  the  melting  behavior, 
specific heat, and  single particle binding energy that  we obtain are 
made  with experiment.   Finally, having  verified our  method  by the 
above comparisons,  we examine the low density,  low temperature phase 
of the  helium monolayer.  We  determine that for all  coverages below 
the  commensurate density, the  system consists  of solid  clusters in 
phase coexistence  with coalesced  vacancies.  No liquid  phase occurs 
and so there is no possibility for first layer superfluidity.

 
\section{Simulation method} 
\label{sec:method} 
 
Our study employs a path-integral Monte Carlo (PIMC) method for simulating 
bose systems.  Details on the application  
of this method to bulk helium can be found 
in the review by Ceperley,\cite{ceprev} and our modifications for the 
simulation of adsorbed helium films may be found in a previous  
publication.\cite{pierce99a} We briefly review the procedure now 
in order to explain the calculations presented in this paper. 
 
PIMC evaluates properties of an $N$-body quantum system at  
the inverse temperature $\tau$ by sampling the partition function  
$Z$.  $Z$ is expanded as a path-integral  
by inserting $M$ intermediate configurations: 
\begin{eqnarray} 
Z \ = & & \frac{1}{N!}  
	\sum_P \int...\int d^3 R_1 ... d^3 R_M d^3 R \nonumber \\ 
  & & \times \rho ({\bf R}_1,{\bf R}_2; \tau) \rho({\bf R}_2, {\bf R}_3; \tau)  
  	\ldots \rho ({\bf R}_M,P {\bf R}_1; \tau), 
\label{eq:pathint} 
\end{eqnarray} 
where $\rho$ is the density matrix, ${\bf R}_i$ is a configuration of  
$N$ particles, and $\tau=\beta/M$.  By taking $M$ large enough, the 
density matrices at $\tau$ can be accurately approximated.  The particular 
PIMC method that we employ ergodically samples both particle positions 
and particle permutations. 
 
The key element needed in the first layer simulation is an accurate 
approximation for the high temperature density matrices.  For bulk 
helium and helium on a smooth substrate, the starting approximation 
for the density matrix can be taken with a temperature as low as 
40 K.  In this paper, results for the first layer on a smooth 
substrate were obtained using the same starting approximation 
for the density matrix that was used in previous calculations 
for the second layer.\cite{pierce99a} 
 
However, as discussed in our previous publication,\cite{pierce99a} 
a helium-graphite interaction that includes substrate corrugations  
makes the starting approximation used for smooth substrates impractical, 
and so we must use a simpler form. 
For these calculations, 
we instead use the high temperature 
expansion of the density matrix.  At sufficiently small values of the 
inverse temperature, the density matrix is given by 
the Trotter approximation: 
\begin{eqnarray} 
\label{eq:trotter} 
\exp(-\tau \hat{H}) \approx \exp(-\tau \hat{T}) \exp(-\tau \hat{V}), 
\end{eqnarray} 
where $\hat{H}$, $\hat{T}$ and $\hat{V}$ are the Hamiltonian and the  
kinetic and potential energy operators, respectively, and 
$\tau$ is the inverse temperature.  This allows us 
to approximate the density matrices at $\tau$ as 
\begin{eqnarray} 
\label{eq:hitemp} 
\rho({\bf R},{\bf R}';\tau) \propto & &\exp[-\pi ({\bf R}_i 
	-{\bf R}_{i+1})/\lambda_T^2 \nonumber \\ 
        & & -\tau (V({\bf R}_i)+V({\bf R}_{i+1}))/2], 
\end{eqnarray} 
where $R$ and $R'$ are the position vectors for $N$ particle configurations,  
and $\lambda_T$ is the thermal wavelength.  This is sometimes referred 
to as the semiclassical approximation, and is accurate for sufficiently 
high starting temperatures. 
Averaging over the potential energy terms is referred to as the 
endpoint approximation.  The potential energy term $V({\bf R}_i)$ is 
the sum of all helium-helium and helium-graphite interactions. 
The exponent of Eq.\ \ref{eq:hitemp} is the first term in an expansion 
in powers of $\tau$.\cite{imada84} 
 
In the semiclassical calculations, we have used 200 K as the starting  
temperature, meaning that 200 inverse-temperature slices are  
required to reach 1 K.  We have verified that this temperature 
is sufficiently high for the approximation to be accurate by comparing 
the energy calculated at 4 K using 200 K and 320 K as the starting  
temperatures.  The energy values obtained agreed within error bars.  We have 
also adopted a three-level bisection, $l=3$, for sampling the positions. 
We verified that this level gave a lower energy at 4 K than 
calculations using $l=2$ and $l=4$.  The acceptance rate using 
$l=3$ is approximately 40\%.  We have further verified that the  
density matrix is well approximated at 200 K by Eq.\ \ref{eq:hitemp} by 
including the next term in the series expansion in powers of $\tau$.   
Energy values calculated with this starting approximation agreed within error 
bars with those that used Eq.\ \ref{eq:hitemp}.   
 
We did find that higher order terms in $\tau$ must be included in the 
energy estimator.  The values we report in Sec.\ \ref{sec:sqr3} are 
obtained using  
\begin{eqnarray} 
\label{engest} 
<E> = \frac{3 N}{2 \tau} + <T> + <V> + \tau^2 \hbar^2/(8 m) <(\nabla V)^2> 
\end{eqnarray} 
for the energy expectation value, 
where $N$ is the number of particles, $<T>$ and $<V>$ are the kinetic and 
potential energy expectation values, and $m$ is the mass of the helium 
atom.  Note that in PIMC the effective action, defined as 
the natural log of the density matrix, not the energy estimator, is 
used to choose between configurations in the Monte Carlo procedure. 
 
Once the machinery for the Monte Carlo method is in place, helium-helium 
and helium-substrate potentials are required for input.  We use 
the Aziz\cite{aziz92} potential for the helium-helium interaction. 
For the helium-graphite interaction, we have adopted the  
anisotropic Lennard-Jones potential proposed by 
Carlos and Cole.\cite{cole80}  This 
potential can be expressed as a Fourier series in the reciprocal lattice 
vectors, {\bf G}, of the graphite substrate.  In cylindrical coordinates 
($\rho$,z) the expansion is 
\begin{eqnarray} 
\label{eq:vgr} 
V({\bf r}) = V_0(z) + \sum_{{\bf G}} V_{{\bf G}}(z) \exp(i{\bf G  
		\cdot \rho}),  
\end{eqnarray} 
where $V_0(z)$ is the laterally averaged potential, and $V_{{\bf G}}(z)$ 
gives the corrugation strength.  The mathematical forms for these 
terms are given elsewhere, \cite{cole80} and the series 
converges rapidly.  For the smooth substrate, we 
use only $V_0(z)$.  For calculations that included corrugations, we 
kept the six lowest, equivalent values of {\bf G} in the expansion.  
 
The limitation of our method is thus related to the accuracy of the 
potentials available to us.  It is possible, for instance, that 
the substrate may substantially mediate the helium-helium  
interaction.\cite{bruch82} 
This is the so-called McLachlan interaction \cite{mclachlan64}.  We have 
performed calculations both with and without this term.  Another possible 
concern is the helium-graphite potential, Eq.\ \ref{eq:vgr}, which 
may overestimate the corrugation strength.  Lowering the corrugation 
height will effect the properties of the first layer, with the favored 
phase becoming liquid instead of solid at sufficiently small corrugations. 
We have examined this effect by repeating some of the calculations with 
the corrugation strength set to 50 \% and to 75 \% of the value obtained from 
the Carlos-Cole model.  Details of these tests of the limits of our 
interaction model are given below. 
 
Particle permutations were also included in the calculations and were 
observed in the film on the smooth substrate at low densities.   
However, we have found that permutations do not play a role in the  
first layer on the corrugated substrate.  We have allowed for  
permutations at intermediate densities for temperatures as low as 
0.571 K with the level of the path bisectioning taken as large as 
5 (32 beads updated in one move) but never observed any particle exchanges.   
This has been checked in calculations both with and without the McLachlan 
interaction.

\section{RESULTS FOR SMOOTH GRAPHITE SURFACE}
\label{sec:layr1} 
 
\subsection{Energy Calculations} 
In this section we report simulation results for the first layer 
obtained by using only the 
laterally averaged portion $V_0(z)$ of the helium-graphite potential. 
We ignore substrate corrugations.   
This system has recently been studied with the Green's Function Monte Carlo 
(GFMC) method,\cite{whitlock98} so we have the opportunity to verify 
that the full implementation of our method is in agreement with 
the results of these complementary calculations. 
The GFMC calculations employ the same helium-graphite potential as we do,  
but use an older form of the helium-helium potential.\cite{aziz79} 
The older form was used since the authors wanted to make direct comparisons 
with previous work on two-dimensional helium using this 
potential.\cite{whitlock88}  The newer potential that  
we use for the helium-helium 
interaction is somewhat more attractive (the well depth is about 0.1 K 
greater), so we expect the energy per particle to be somewhat lower 
near the equilibrium density.  This has 
been observed in recent zero-temperature calculations for two dimensional    
liquid helium that employ the newer potential.\cite{giorgini96}  
 
Guided by previous simulations, we expect the helium film to have 
a self-bound liquid phase and to solidify at high densities, before 
promotion to the next layer occurs.  Calculations at low and intermediate 
densities are performed using a square simulation cell, while solid phase 
calculations use a rectangular simulation cell that accommodates 
the periodic structure of the triangular solid.   
It is not necessary in PIMC to employ different 
forms for the density matrix for the liquid and solid phases.   
Liquid calculations employed  
16 particles, while solid phase calculations used 30 particles.  Both sets of  
calculations were performed at 400 mK.  Calculations at 500 mK are 
in agreement with these results, indicating that we have converged to the 
zero-temperature limit.  We verified that there were no finite-size 
errors in the liquid phase by repeating some of the calculations  
with 32 particles.  The energy values at the two temperatures were 
in agreement.  Finite size errors were 
found to be negligible for the solid phase also, since the energy value 
calculated at 0.0689 atom/$\AA^2$ using the rectangular  
simulation cell (30 particles) agreed with the value  
calculated using the square simulation cell (16 particles).   
 
These calculations are somewhat different from those we discuss in 
the other section of this paper. 
The potential between the active helium  
atoms and the underlying substrate 
is featureless in the plane of the substrate, so the size of the  
simulation cell may  
be varied continuously.  This allows us to keep the number of particles  
constant for each phase.  In contrast, other calculations we have performed 
(including those with substrate corrugations taken into account)  
were with a varying number of particles and a constant simulation cell size.   
The two methods lead to somewhat different forms for the Maxwell  
construction.  In the present case, liquid-solid coexistence will be 
characterized by a linear region in the dependence of the energy per 
particle on the atomic area (inverse volume). 
 
\begin{figure}[htp] 
\epsfxsize=\figwidth\centerline{\epsffile{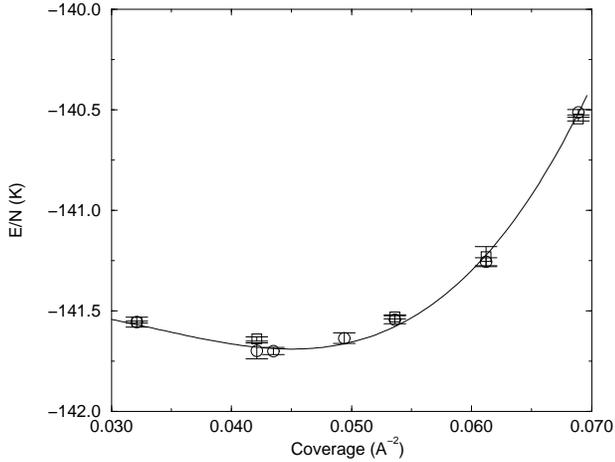}} 
\caption{Energy values for the first layer liquid.  The circles are 
our results obtained with PIMC.  The squares are  
GFMC results.\cite{whitlock98}  The line is a least squares fit  
of the polynomial, Eq.\ (\ref{eq:polyl}) to our data. 
} 
\label{fig:engliq} 
\end{figure}

\begin{figure}[htp] 
\epsfxsize=\figwidth\centerline{\epsffile{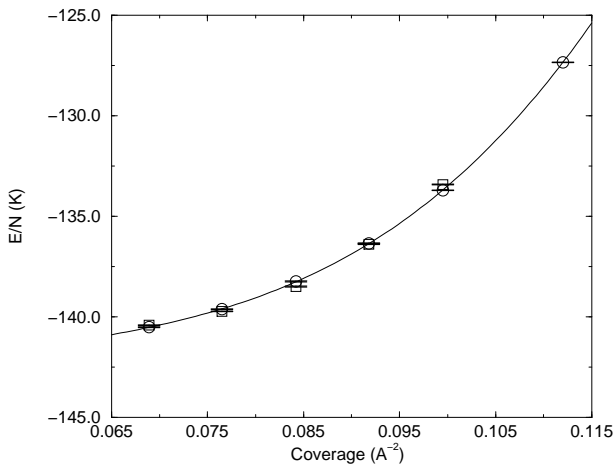}} 
\caption{Energy values for the first layer triangular solid.  The 
symbols have the same meaning as in Fig.\ \ref{fig:engsolid}.  The line 
is determined from the fit of Eq.\ (\ref{eq:polys}) to our data. 
} 
\label{fig:engsolid} 
\end{figure}

Figure \ref{fig:engliq} displays our results for the energy of the 
first layer liquid.  Also shown are the results obtained from GFMC. 
For all points except the near equilibrium, the two calculations  
agree within error bars.  Near the energy minimum, there is some 
disagreement, but even here the computed values differ only by about 0.6\%. 
This is perhaps attributable to the different helium-helium potentials 
used in the two calculations.  Figure \ref{fig:engsolid} 
shows our results and the GFMC results for the first layer solid 
phase.  Again, there is excellent agreement between the two methods.  
 
Following the typical procedure, we have determined the equations of state 
for the liquid and solid phases by fitting our energy values to 
the polynomials 
\begin{eqnarray} 
&& E/N=e_0 + B(\frac{\rho-\rho_0}{\rho_0})^2 + C(\frac{\rho-\rho_0}{\rho_0})^3,\\ 
\label{eq:polyl} 
&& E/N=\alpha + \beta \rho + \gamma \rho^2 + \delta \rho^3. 
\label{eq:polys} 
\end{eqnarray} 
for the liquid and solid phases, respectively.  Values for the  
fitted parameters are given in Table \ref{tab:fit}.  Errors for the liquid 
phase are indicated.  The values shown for the solid phase were used in 
the plot of Fig.\ \ref{fig:engsolid}.  The values given for 
$\beta$,$\gamma$, and $\delta$ are accurate to one significant figure. 
 
For the liquid phase, the  
parameters $e_0$ and $\rho_0$ are the equilibrium energy per particle and the 
equilibrium coverage.  The equilibrium density  
we obtain, 0.0450 atom/$\AA^2$, is 
in reasonable agreement with the GFMC value of 0.0443.  Below this density, 
the system enters gas-liquid coexistence in the thermodynamic limit. 
In simulation, phase separation will not occur immediately because of the 
finite cost of creating the phase boundary.  The system will instead enter 
a ``stretched'', or negative pressure, state.  If one continues to decrease 
the coverage, however, the phase separated state will eventually become  
the favored state, and the ``stretched'' state will ``snap'', forming 
a droplet plus vacuum.  For a simulation with a constant number of  
particles and variable area, this occurs when the derivative 
of the spreading pressure, $P=\rho^2(\partial e(\rho)/\partial \rho)$, is 
zero.  At the spinodal point, the sound velocity becomes imaginary and 
the compressibility diverges. 
From the equation of state, we determine that this occurs  
at 0.034 atom/$\AA^2$.  For comparison, the spinodal density has been 
calculated to be between 0.031 and 0.038 atom/$\AA^2$ for two-dimensional 
helium.\cite{clements93b,giorgini96,whitlock88}   
 
\begin{table} 
\begin{center} 
\begin{tabular}{|l r | l r|} \hline 
Parameter          & Liquid       & Parameter           & Solid  \\ \hline 
e$_0$(K)           & -141.689(14) & $\alpha$(K)         & -157.46    \\ 
$\rho_0$($\AA^{-2}$)  & 0.0450(06)  & $\beta$(K/$\AA^2$)  & -679.83    \\ 
B(K)               & 2.42(27)     & $\gamma$(K/$\AA^4$) & -10504.77  \\ 
C(K)		   & 3.29(67)     & $\delta$(K/$\AA^6$)	& 61032.24   \\ 
$\chi^2/\nu$       & 2.57         & $\chi^2/\nu$        & 1.3   \\ \hline 
\end{tabular} 
\caption{ Fitted parameters for the liquid and solid equations of state.  The 
errors in the last digits of the parameters for the liquid phase are given in  
parenthesis. 
} 
\label{tab:fit} 
\end{center} 
\end{table} 
 
\begin{figure}[htp] 
\epsfxsize=\figwidth\centerline{\epsffile{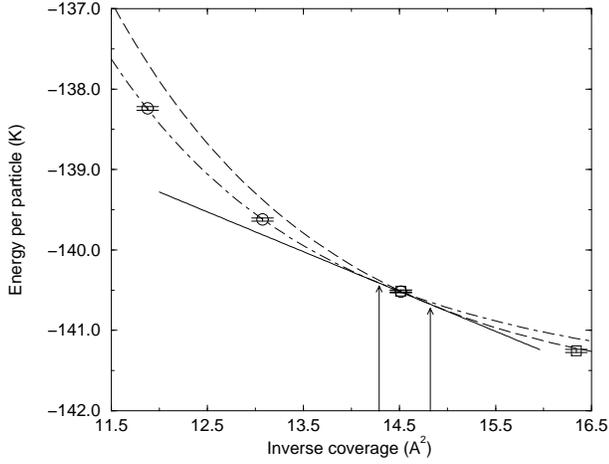}} 
\caption{Liquid-solid coexistence regions determined with the  
Maxwell construction.  The dashed-dotted and dashed lines are the  
solid and liquid  
equations of state, respectively.  Circles with error bars are  
calculated energy values for the liquid.  Squares with error bars show 
values calculated for the solid phase. 
The unbroken line is the coexistence line. 
The arrows indicate the beginning and end of the coexistence region.  
} 
\label{fig:maxwell} 
\end{figure}

From the polynomial fits for the two layers, we can estimate the regions 
of phase coexistence for the two solids by using the Maxwell double 
tangent construction.  See Fig.\ \ref{fig:maxwell}. 
Since we have a constant number of particles and a variable density, the 
Maxwell construction is found from the common tangent of the energy per 
particle versus atomic area (inverse coverage).  We determine the 
liquid-solid coexistence region to be from 0.0675 to 0.0700 atom/$\AA^2$. 
This is comparable to the ranges found for both two-dimensional  
helium\cite{whitlock88,gordillo98} and  
the helium film,\cite{whitlock98} although 
the onset of full solidification that we find is at a somewhat lower density.   
The solidification density determined in this manner is subject to  
error, since our fitted parameters are not determined with a high 
degree of precision.   
 
\begin{figure}[htp] 
\epsfxsize=\figwidth\centerline{\epsffile{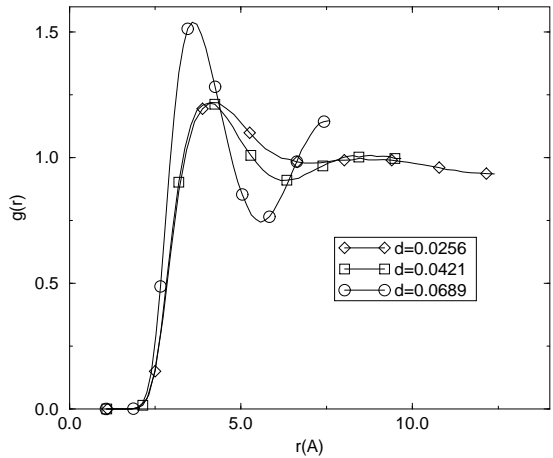}} 
\caption{Radial distribution function for the liquid phase at 
the indicated densities, in atom/$\AA^2$. 
} 
\label{fig:rdflayr1} 
\end{figure} 
 
\begin{figure}[htp] 
\epsfxsize=\figwidth\centerline{\epsffile{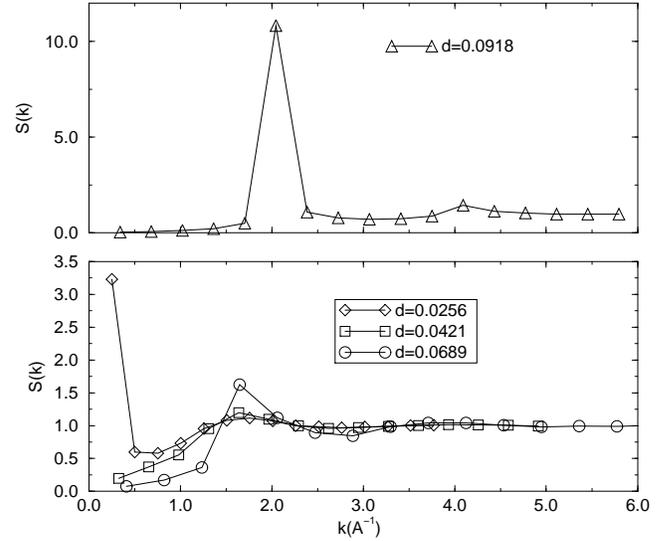}} 
\caption{Static structure function for the liquid phase at 
the indicated densities, in atom/$\AA^2$.  The static structure function 
in the (01) direction at a typical solid density (top) is also shown. 
} 
\label{fig:ssflayr1} 
\end{figure} 
\begin{figure}[htp] 
\epsfxsize=\figwidth\centerline{\epsffile{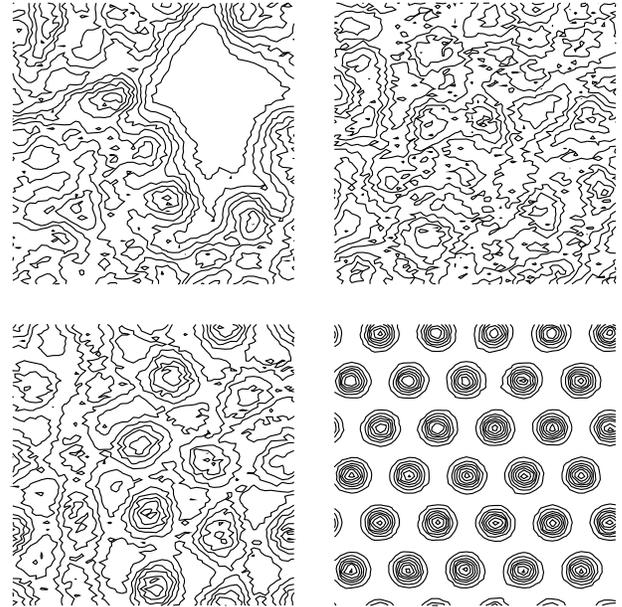}} 
\caption{Contour plots of the probability density in the plane of the 
substrate.  The densities shown are, top row, left to right, 0.0256 
and 0.0421 atom/$\AA^2$; bottom row, left to right, 
0.0689, and 0.0918 atom/$\AA^2$. 
} 
\label{fig:contour1} 
\end{figure} 
\begin{figure}[htp] 
\epsfxsize=\figwidth\centerline{\epsffile{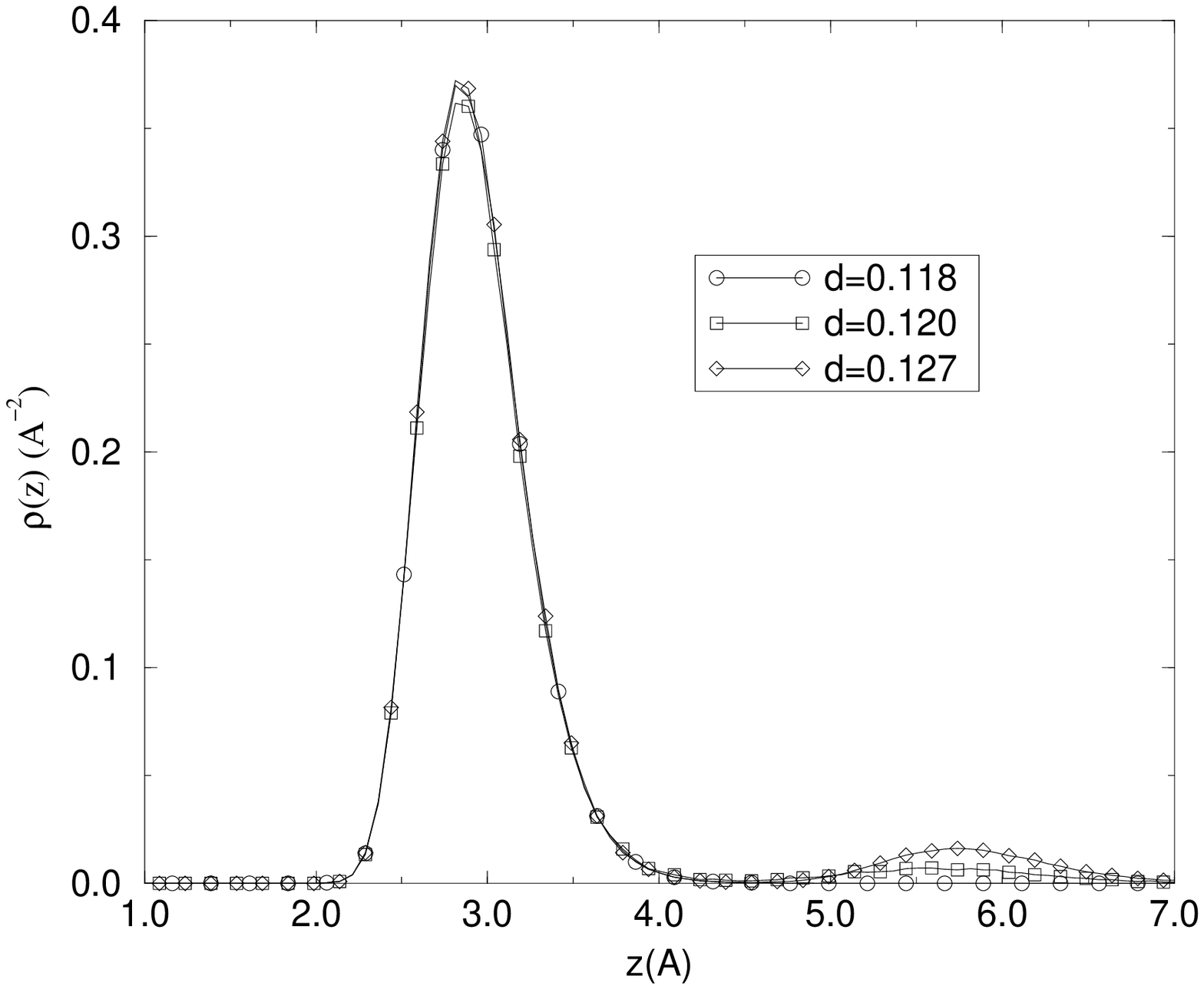}} 
\caption{Density profiles near second layer promotion.  The coverages, in 
atom/$\AA^2$, are indicated.  Normalization is chosen so that integrating 
the profiles gives the coverage. 
} 
\label{fig:zdistlayr1} 
\end{figure} 
 
\subsection{Other properties} 
As we have seen, in the absence of corrugations, the first layer 
has liquid-gas, self-bound liquid, liquid-solid,  
and solid regions.  The  
gas phase has zero density at zero temperature.   
We expect the liquid-gas phase to have two 
regions, one in which phase boundaries between these phases form, and 
one in which the liquid phase is artificially stretched.  This is 
governed by the energy cost required to create the phase boundary.  The 
metastable liquid phase is spatially indistinguishable from the  
equilibrium state, but the formation of droplets can be visualized. 
In Fig.\ \ref{fig:rdflayr1} we plot the radial distribution function g(r)  
for the liquid at coverages in the droplet region, near equilibrium, and near 
the onset of liquid-solid coexistence.  Here, $r$ is the magnitude of the 
projection of the 
distance vector between two particles onto the plane of the substrate. 
In the droplet region, the 
long range tail drops below unity, indicating the system does not uniformly 
cover the substrate.  Near equilibrium the first peak has about the same 
height as the droplet phase, but the tail goes to unity at long ranges.  The 
peaks are located at the same value of $r$, indicating that the average 
separation 
distance of nearest neighbor particles in the droplet is  
close to the equilibrium value. 
The high density liquid has noticeably more correlation.  The peak height is 
1.5, compared to 1.2 near equilibrium, and the peak's position has shifted from 
4.2 to 3.6 $\AA$, as one would expect from compression caused by increasing 
the density.  These values are in agreement with values reported for 
the helium liquid in two dimensions,\cite{whitlock88} indicating that the 
first layer liquid is very two-dimensional in character. 
 
The results for the static structure factor $S(k)$ for the liquid phase are 
shown in Fig.\ \ref{fig:ssflayr1} and can be interpreted similarly.  At the 
lowest coverage, $S(k)$ swings upward for low values of $k$ instead of  
going to zero.  This indicates the presence 
of droplets and a nonzero compressibility. 
For the two uniform liquid coverages,  
$S(k \rightarrow 0) \rightarrow 0$.  The structure function 
is more peaked for the high density liquid.  Figure \ref{fig:ssflayr1} also 
shows results for a typical solid coverage.  The peak heights for the  
fluid near equilibrium and for the dense fluid are 1.2 and 1.6, respectively. 
These values are in reasonable agreement with, but somewhat below, the 
two-dimensional values.

Finally, we show in Fig.\ \ref{fig:contour1} probability contours for the 
first layer for the various regions discussed above.  The lowest 
coverage shown is below the spinodal point, and as expected  
incompletely covers the substrate.  Near the equilibrium 
density, the substrate is uniformly covered.  In the dense liquid, localization 
can be observed.  This coverage is in the liquid-solid coexistence region. 
Finally the system enters a triangular solid phase at 
the highest density.

Promotion to the second layer may be determined by examining the 
density profiles for the dense solid.  These are shown in  
Fig.\ \ref{fig:zdistlayr1} at the indicated coverages.  The occupation of 
the second layer is clearly visible at the coverages 0.1200 and 0.1270 
atom/$\AA^2$, while the coverage 0.1180 shows no evidence of promotion. 
Integrating the profiles for the two largest coverages to the minimum 
between the peaks (4.5 $\AA$) gives 0.116 and 0.119 atom/$\AA^2$ for 
the first layer coverage.  Experimentally,  
estimates\cite{bretz73,bretz78,carneiro81,greywall91} of first layer 
completion range from 0.112 to 0.120 atom/$\AA^2$ 
so our value is in good agreement.  The recent GFMC  
calculations\cite{whitlock98} obtain a value between 0.115 and 0.118 for 
the beginning of second layer promotion. 

\section{RESULTS WITH CORRUGATED SUBSTRATE}
\label{sec:sqr3} 
 
In this section we present results obtained by using a more 
realistic model of the graphite substrate.  Most of the  
results shown in this section were obtained with a simulation 
cell with the dimensions $25.56\AA \times 22.14 \AA$.  The commensurate 
density corresponds to 36 helium atoms.  Periodic boundary 
conditions were used in the plane of the substrate.  Finite-size effects 
were examined by repeating some calculations using a $34.08 \AA \times 
29.51 \AA$ simulation cell, for which the commensurate density  
corresponds to 64 particles. 
 
\subsection{At And Below Commensurate Density} 
 
\begin{figure}[ht] 
\epsfxsize=\figwidth\centerline{\epsffile{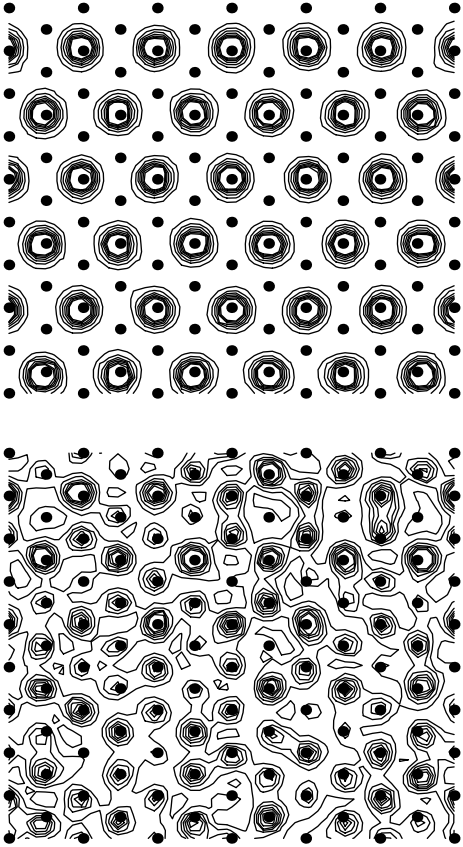}} 
\caption{Distribution plot at the commensurate density, 0.0636 atom/$\AA^2$, 
for  T=2.99 K and T=4.0 K.  The filled circles indicate graphite 
adsorption sites.} 
\label{fig:dist636} 
\end{figure} 
 
The principal conclusion we obtain from these calculations is that the low 
density, low temperature phase of the first layer consists of commensurate solid 
clusters, rather than a liquid phase.  In  an earlier publication\cite{pierce99b}  
we observed the commensurate phase and investigated its melting 
with both static structure and specific heat calculations. 
The calculated $\sqrt 3 \times \sqrt 3$ commensurate solid phase  
and its melt are shown in Fig.\ \ref{fig:dist636}.  At 3 K, 
the film has solidified into the commensurate structure.   
The solid forms a sub-lattice that contains 
one-third of the adsorption sites.  The remaining two-thirds of the sites 
form two equivalent sub-lattices that are unoccupied. 
Raising the temperature to 4 K causes  
melting.  At this temperature the helium atoms no longer inhabit a single 
sub-lattice of substrate adsorption sites.  All adsorption sites will be 
occupied with equal probability if the simulation is run for a sufficiently 
long time. 
 
\begin{figure}[ht] 
\epsfxsize=\figwidth\centerline{\epsffile{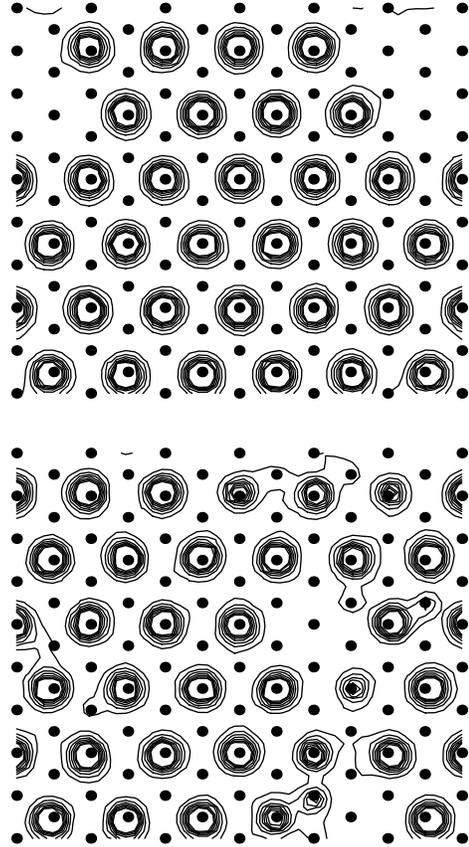}} 
\caption{Probability distributions for 0.0566 atom/$\AA^2$ at 
T=1.0 (top) and 2.5 K (bottom). 
The filled circles give the locations of graphite potential minima. 
} 
\label{fig:distrib566} 
\end{figure} 
 
\begin{figure}[ht] 
\epsfxsize=\figwidth\centerline{\epsffile{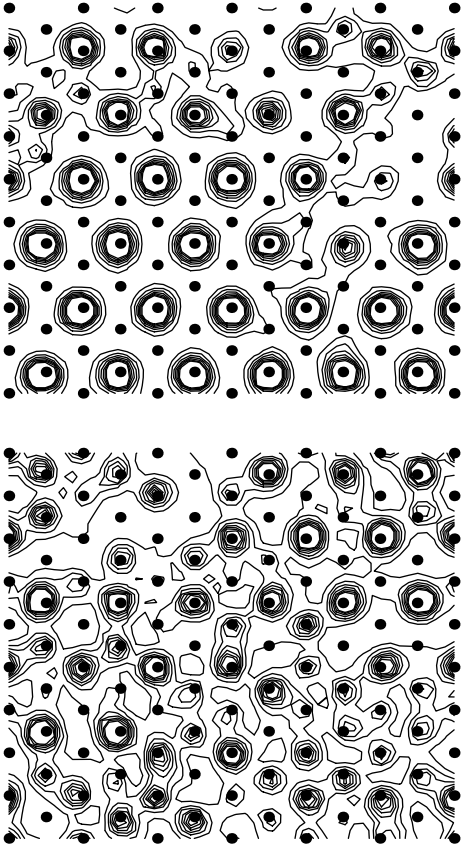}} 
\caption{Distribution plot at 0.0530 atom/$\AA^2$, T=2.5 K  
and T=3.0 K} 
\label{fig:dist530} 
\end{figure} 
 
\begin{figure}[ht] 
\epsfxsize=\figwidth\centerline{\epsffile{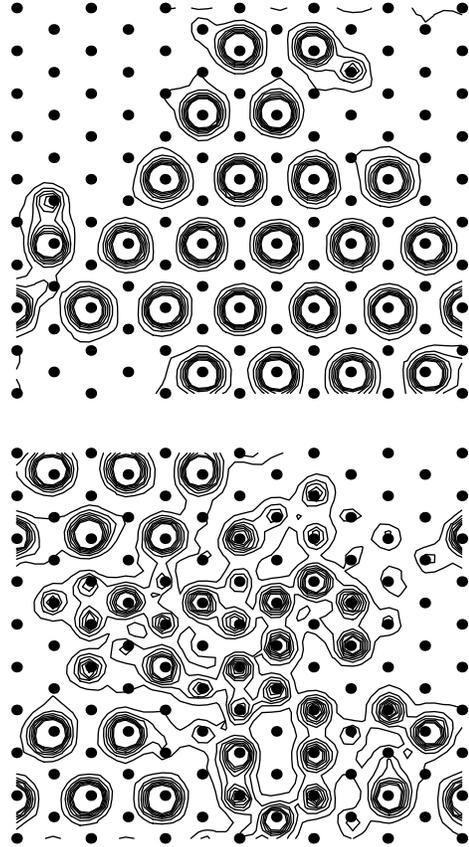}} 
\caption{Distribution plot at 0.0424 atom/$\AA^2$, T=1.0 K  
and T=2.0 K} 
\label{fig:dist424} 
\end{figure} 
 
\begin{table} 
\begin{center} 
\caption{Estimates of melting temperatures from the temperature 
dependence of the static structure peaks.} 
\vspace{.2in} 
\begin{tabular}{|c|c|} \hline 
Coverage (atom/$\AA^2$) & T$_{melt}$ \\ \hline 
0.0424 & 2.0 \\ 
0.0530 & 2.5 \\ 
0.0566 & 3.0 \\ 
0.0636 & 3.33 \\ \hline 
\end{tabular} 
\label{tab:melt} 
\end{center} 
\end{table} 
 
Next, we present direct evidence that the low density (below the commensurate solid  
density of 0.0634 atom/$\AA^2$) first layer consists of a solid cluster surrounded  
by a low density vapor at low temperatures. No liquid phase forms and so there  
is no possibility for first layer superfluidity.  These findings are in contrast  
to the most recent proposal for this region\cite{greywall91}. 
 
The presence of a solid with vacancies and phase separation can be visualized 
with contour plots of the probability distributions, shown  
in Fig.\ \ref{fig:distrib566} for the indicated coverage and 
temperatures.  At the lower temperature, 1.0 K, the vacancies have condensed 
into a single bubble region, as can be seen in the top plot of  
Fig.\ \ref{fig:distrib566}.  Note that periodic boundary conditions are 
being used, so the vacancy regions in the figure are  
connected.  The vacancies move very slowly at this temperature; we 
found equilibration times to be very long for the vacancies to 
condense if they were initially  
spread throughout the lattice.  We thus calculated the 
energy for this system twice:  first with the vacancies initially spread  
through the solid and then with the  
vacancies condensed.  The condensed energy was found to be lower.   
At higher temperatures, the vacancies acquire enough kinetic energy 
to leave the phase separated state and diffuse into the solid.   
As a result, vacancies can 
become isolated.  This is illustrated at 2.5 K in the lower plot 
of Fig.\ \ref{fig:distrib566}.  A series of probability distribution plots  
reveals that these vacancies move in the simulation, so the equilibration 
problem encountered at 1.0 K is not present at this temperature. 
We note that we still see evidence of phase separation in contour plots 
at 2.0 K, while heat capacity peaks from experiment seem to indicate 
a transition at 1.5 K.

Figures \ref{fig:dist530} and \ref{fig:dist424} further confirm 
that the low density region contains solid clusters at low 
temperatures and that these clusters exhibit 
the melting behavior discussed in the previous section.  We have calculated 
distribution plots for densities as low as 0.0207 atom/$\AA^2$ and observe 
solid clusters at all densities. 
 
We have also attempted to place a vacancy in a solid cluster to see if 
the cluster could support an isolated vacancy and at the same 
time be in equilibrium with the low density vapor. 
We found that at 0.0424 atom/$\AA^2$ and 1.0 K, 
the vacancy was spontaneously expelled from the solid cluster 
during thermalization.  We conclude that  
at low temperatures the solid clusters will not support isolated vacancies.

Further evidence for solidification into the commensurate 
structure in the simulation comes from calculations of the  
static structure factor.  Typical results for static structure  
factors for coverages at and immediately below the 
commensurate solid density have been reported in our earlier paper\cite{pierce99b}. 
In addition, the melting of the commensurate solid phase can be  
determined from the temperature dependence of the static structure  
peak height. Melting is signaled by a significant drop in the peak  
height and a large statistical fluctuation in a peak  value near  
the melting temperatures.  The melting temperature determined in this  
manner are given in Table \ref{tab:melt}. The density dependence  
of the melting temperature is consistent with the experimental  
phase diagram, although our melting temperatures are slightly higher than the  
experimental values.  Heat capacity measurements indicate that  
the commensurate solid melts at about 3 K, and the low density  
($< 0.045$ atom/$\AA^2$) melting peaks are at about 1.5 K. 
 
\begin{figure}[ht] 
\epsfxsize=\figwidth\centerline{\epsffile{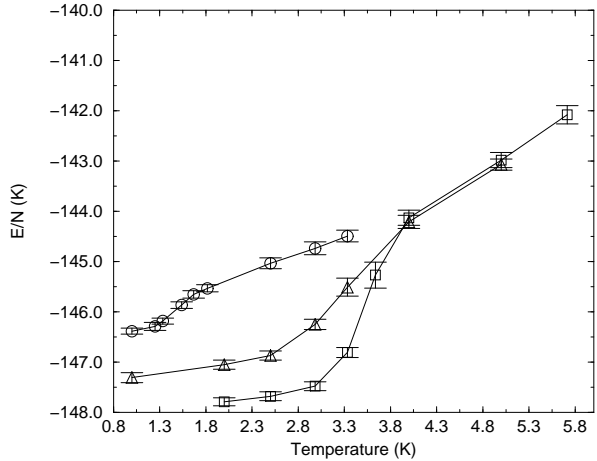}} 
\caption{Temperature dependence of the energy per particle.   
The coverages are 0.0353 (circles), 0.0566 (triangles),  
and 0.0636 atom/$\AA^2$ (squares).} 
\label{fig:engtemp} 
\end{figure} 

\begin{figure}[ht] 
\epsfxsize=\figwidth\centerline{\epsffile{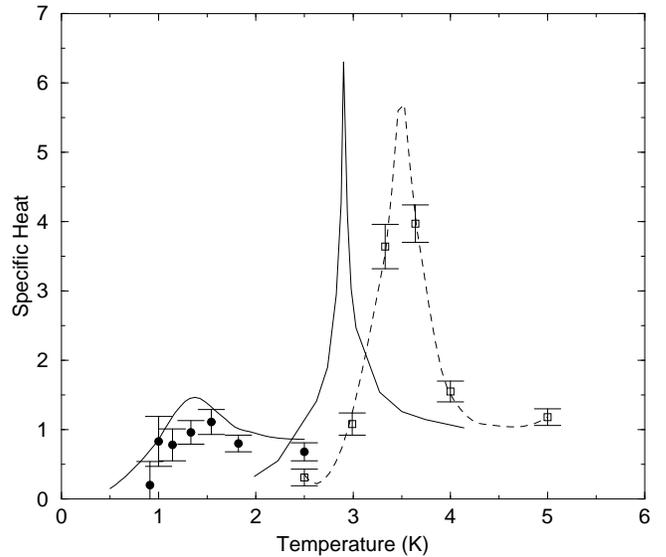}} 
\caption{Specific heat for 0.0353 (filled circles) 
and 0.0636 atom/$\AA^2$ (squares).  The dashed line is a guide to 
the eye.  The solid lines are the experimentally determined specific heat  at 0.0367  
and 0.0634 atom/$\AA^2$. 
} 
\label{fig:specheat} 
\end{figure} 

Another way of estimating melting temperatures is from the  
temperature dependence of the energy. This is shown in  
Fig.\ \ref{fig:engtemp} for various densities.   
These curves possess inflection points that lead to specific 
heat maxima when differentiated.  These peaks 
indicate melting.  Figure \ref{fig:specheat} shows two sample  
calculations of the specific heat at the indicated densities.   
The peak height and location change dramatically with density.  Melting  
occurs at about 1.5 K and 3.5 K for 0.0353 (filled circles) and 0.0636 atom/$\AA^2$ (open squares) 
respectively.  In Fig.\ref{fig:specheat} the calculated specific heat for 0.0353  and  
0.0636 atoms/$\AA^2$  is shown. For comparison we also present the 
experimental measure specific heat (solid lines) for densities of 0.0367 and 0.0634 atoms/$AA^2$. 
The calculated specific heat has peaks  somewhat above the experimental values but 
are consistent with the melting behavior exhibited in the contour plots and in the static structure 
factor calculations.

\subsection{Maxwell Construction} 
 
\begin{figure}[ht] 
\epsfxsize=\figwidth\centerline{\epsffile{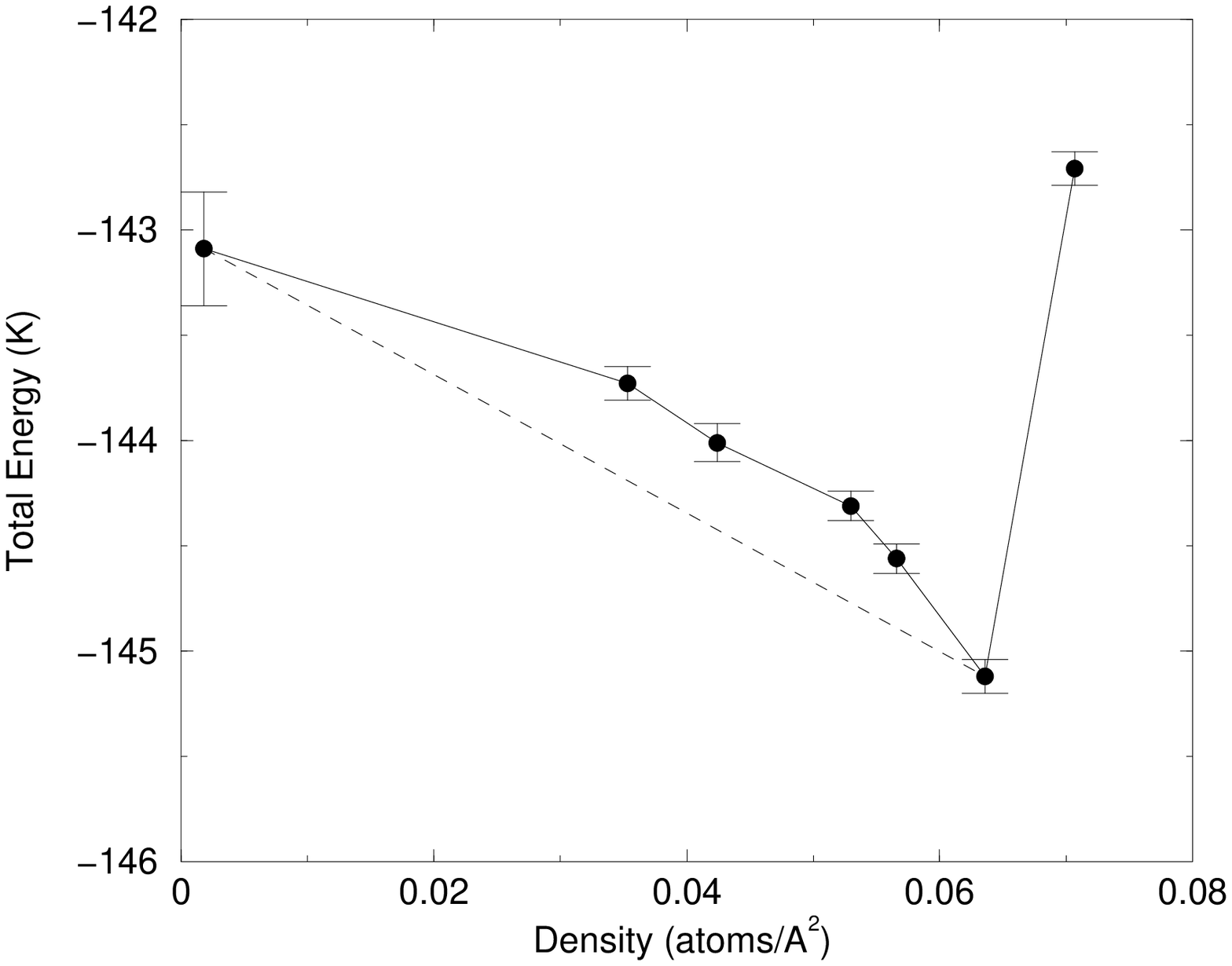}} 
\caption{Total energy versus coverage.  The dashed 
line is a straight line connecting the lowest density accessible in our 
lattice and the density that corresponds to 1/3 coverage. 
} 
\label{fig:engsqr3} 
\end{figure} 
\begin{table} 
\begin{center} 
\caption{Energy/particle versus coverage.  The first column gives the  
temperature of the calculation.  The number in parenthesis gives the 
error in the last decimal place.} 
\vspace{.2in} 
\begin{tabular}{|c|c|c|} \hline 
T(K) & $\sigma$($\AA^{-2}$) & E/N (K) \\ \hline 
1.00 & 0.0353 & -143.73(8) \\ 
1.00 & 0.0424 & -144.03(9) \\ 
1.00 & 0.0530 & -144.31(7) \\ 
1.00 & 0.0566 & -144.56(7) \\  
1.00 & 0.0636 & -145.12(8) \\ 
1.33 & 0.0707 & -142.71(8) \\ \hline 
\end{tabular} 
\label{tab:engsqr3} 
\end{center} 
\end{table} 
 
 
We determine phase ranges by applying the Maxwell construction to the 
total ground state energy.  For a system at constant volume with a  
varying number of particles, a region of phase separation will be 
signaled by an unphysical upward curvature in the total free energy's 
dependence on density.  The upper and lower bounding densities of this 
region are connected by a common tangent line.  The total free energy 
for all intermediate densities lies on or above this line, either because 
creating a surface between the two phases costs a finite amount of  
energy or because the system remains unphysically homogeneous.  In the 
thermodynamic limit the system will separate into the two phases 
at the bounding densities. 
 
The Maxwell construction at nonzero temperatures should be applied to the 
total free energy.  This is not directly accessible from PIMC, however. 
Instead, we use a limiting process to determine effectively ground 
state energy values.  All energy calculations are performed at low 
temperatures.  The temperature is then raised and the energy is  
recalculated.  If the two values are the same within error bars, we 
conclude that we have obtained effectively zero temperature energy values. 
This allows us to apply the Maxwell construction to the total energy, 
since it is the same as the total free energy at zero temperature. 
 
We now can apply this procedure to the total energy calculations of the 
first layer solid.  Total energy calculations  
for a range of densities using a simulation cell  
designed to accommodate the commensurate solid are shown in 
Fig.\ \ref{fig:engsqr3}.  The values for the energy per particle 
are given in Table \ref{tab:engsqr3}.  
The cell dimensions 
are $25.560 \AA \times 22.136 \AA$, and the number of particles varies 
from 20 to 40.    
The dashed line in Fig.\ \ref{fig:engsqr3} is a straight line  
connecting the lowest density accessible in our 
lattice and the density that corresponds to 1/3 coverage. 
Notice that all intermediate energy values are 
above this straight line.  From the Maxwell construction, this 
indicates that the intermediate energy values are unstable and will 
phase separate into the two stable phases (the vapor and the  
commensurate solid) that bound the unstable region. 
 
The binding energy for a single particle on the substrate can be 
easily calculated.  We find this to be 
$E_B=-143.09 \pm 0.27$ .  This is comparable to 
the estimated values \cite{cole81}  
$E_B=-141.75 \pm 1.50$ K from scattering \cite{derry79,carlos80} 
and $E_B=-142.33 \pm 1.97$ K 
from thermodynamic analysis \cite{elgin74}. 
Our value for the binding energy 
was calculated at 0.4 K and confirmed to be the ground 
state value by the limiting procedure discussed  
above.  By subtracting this energy from 
the energy per particle  
of the commensurate solid phase, we obtain the condensation 
energy per particle for the two-dimensional solid, 
$E_{2D}=-2.03 \pm 0.20$ K.  This is comparable to, but slightly higher 
than, the two-dimensional  
energy (-1.06 K) for the commensurate phase  
found by the variational calculations of 
\cite{bruch93} for the same interaction model.

\subsection{Above Commensurate Density} 
 
\begin{figure}[ht] 
\epsfxsize=\figwidth\centerline{\epsffile{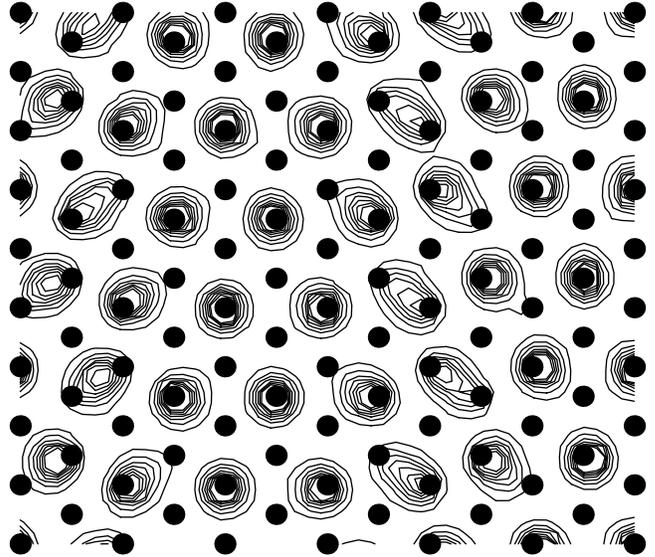}} 
\caption{Distribution plot of the domain wall solid at 0.0742 atom/$\AA^2$ 
(N=42 atoms) and T=1.0 K.  Filled circles indicate adsorption sites. 
} 
\label{fig:dist742} 
\end{figure}

\begin{figure}[ht] 
\epsfxsize=\figwidth\centerline{\epsffile{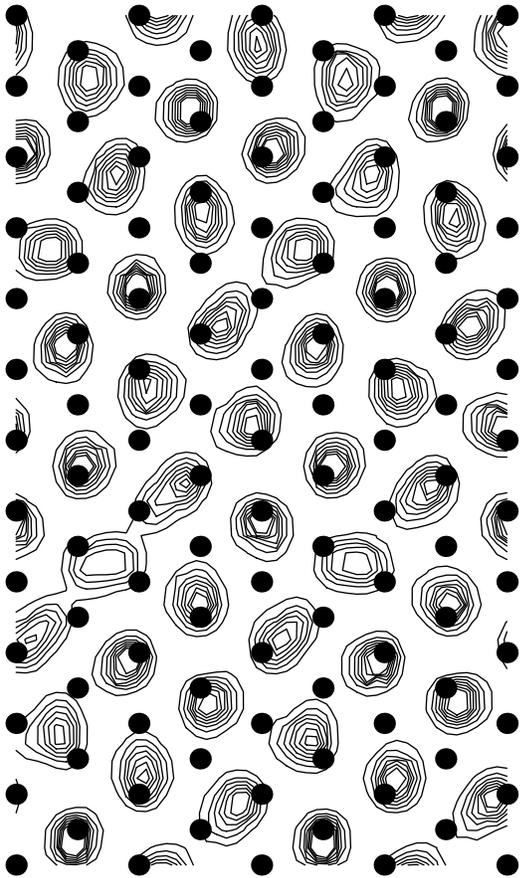}} 
\caption{Distribution plot of the incommensurate solid phase at 0.0994  
atom/$\AA^2$ (N=50 atoms) and T=2.0.  Filled circles indicate adsorption 
sites. 
} 
\label{fig:dist994} 
\end{figure}

Figures \ref{fig:dist742} and \ref{fig:dist994} illustrate the domain 
wall solid and incommensurate triangular solid phases, respectively.  The 
domain wall solid consists of patches of commensurate solid on different 
sub-lattices.  Linear domain walls occur at the boundaries between  
these regions.  In the incommensurate solid, the atoms form a  
triangular solid that does not 
have a periodic relationship with the underlying adsorption sites for 
lengths scales less than the minimum dimension of the simulation cell. 
 
\section{Tests of  Our Conclusions} 
 
The strength of our conclusions is limited by the accuracy of the 
interaction model that we use.  First, it is possible that the 
substrate may substantially alter the helium-helium  
interaction\cite{mclachlan64,bruch82}.  Inclusion 
of this effect, the so-called McLachlan interaction, 
has been shown to change the ground state phase from the  
commensurate solid to a low density liquid in two-dimensional variational 
calculations.\cite{bruch93} We have repeated the  
low temperature density scans 
using the same mediated interactions employed in these 
calculations\cite{bruch93}.  
We found that while the energy per particle increases for 
all coverages, the commensurate solid 
remains the energetically favored phase.  Second,  
another potential problem is that 
the helium-graphite interaction\cite{cole80} is too corrugated, thus favoring 
solidification.  We have repeated the 
low temperature calculations at the commensurate density and at 
0.0424 atom/$\AA^2$ with the corrugation strength reduced by 25 \%.  
The commensurate phase remains energetically favored.  Calculations 
were also performed with the corrugation reduced by 50 \%, but it was  
found that the commensurate solid would not form for temperatures as low 
as 2 K, thus indicating the corrugations had been underestimated.   
We note finally that the melting behavior of the commensurate solid phase 
was not sensitive to the inclusion of the McLachlan term.   
 
Finally, we wish to discuss the arguments of Greywall and Busch (GB) 
against solid clusters and in favor of the superfluid phase. 
Their primary objection to solid-vapor 
coexistence is that this should be signaled by 
linear heat capacity isotherms for the entire region from 
zero coverage up to the commensurate density.  Their 
published data shows that for temperatures from 0.2 K to 
0.5 K, the isotherms are linear only between 0.025 and 0.060 atom/$\AA^2$.   
At 0.1 K, the upper endpoint is about 0.055 atom/$\AA^2$.   
As a possible explanation, we suggest that the departure from linearity below 
0.025 atom/$\AA^2$  is caused by the presence of multiple finite-sized clusters.  At low densities, 
solid clusters nucleate around surface defects. 
Initially, there are many small metastable clusters with large  
perimeter-to-area ratios.  Increasing the density increases the size of the clusters until 
the surface is covered by a few large solid clusters with negligible 
boundary effects.  Thus, the heat capacity exhibits linear  
behavior only after the solid clusters are sufficiently large 
so that the perimeter-to-area ratio is small.  This 
presumably occurs for coverages above 0.025 atom/$\AA^2$.   
GB have used a similar explanation 
in their arguments for solid-liquid and liquid-gas coexistence in  
regions that do not have linear isotherms. 
 
GB's identification of coverages near 0.04 atom/$\AA^2$ as 
liquid is based partly on simulation results for 2D helium on a flat 
substrate, the most relevant calculations then 
available.  As GB note, the large peak associated with  
the melting of the uniform commensurate solid phase first  
emerges above 0.04 atom/$\AA^2$.  2D helium is a liquid 
near this density \cite{whitlock88}, suggesting that  
first layer coverages below 0.04  
may be liquid.  Unlike the purely 2D simulations,  
our calculations take the role of 
substrate effects into account.  As we have shown, surface corrugations 
push the density of the  
energy minimum up from about 0.04 on a flat substrate to 0.0636 
atom/$\AA^2$ and produce solidification.   
GB also show that their low density heat capacity 
results are in general agreement with a PIMC  
calculation for 2D superfluid helium \cite{ceperley89},  
suggesting that there might be a superfluid 
transition in the first layer.  We have 
shown in Fig. \ref{fig:specheat}  
that these rounded heat capacities  
are produced by the melting of a solid cluster and are not associated with  
a superfluid transition. 
 
In closing, we would like to remark that a promising direction for  
monolayer superfluidity\cite{greywall91,chen94,boninsegni98,cowan98} lies  
with helium adsorbed on alkali substrates, 
particularly lithium \cite{cheng89,boninsegni98}. 
These substrates have much smaller corrugations and 
a much weaker attraction, allowing the first layer helium film to be a  
liquid.  The phenomenon competing with superfluidity for these substrates 
is pre-wetting, rather than solidification.

\section{Summary} 
Our first layer calculations were performed both with and without  
substrate corrugations.  When corrugations are neglected, 
the first layer film resembles a purely two-dimensional film.  We determined 
that the film consists of gas, liquid, and solid phases.  These are separated 
by coexistence regions, and we determined the coverage ranges for all 
phases at low temperatures by using the Maxwell construction.  The first 
layer liquid has an equilibrium density of 0.0450 atoms/$\AA^2$.  Below this 
density the system phase separates.  This region is divided into an 
unstable region, where liquid droplets form, and a metastable region in 
which the system over-expands instead of forming an interface.  These 
two regions are separated by a spinodal point at 0.034 atoms/$\AA^2$.  At 
higher densities the system enters a narrow region of liquid-solid coexistence 
between 0.0675 and 0.0700 atoms/$\AA^2$.  Above these coverages the system 
is in a triangular solid phase.  The beginning of layer promotion occurs 
between 0.116 and 0.119 atoms/$\AA^2$.  All of our calculations are in 
agreement with recent Green's function Monte Carlo results.\cite{whitlock98} 
 
When corrugations are included in the first layer, the phase diagram is 
substantially altered.  We find that a $\sqrt 3 \times \sqrt 3$ commensurate 
solid occurs, in agreement with numerous experiments.  By examining  
the temperature dependence of the static structure function and the  
specific heat, we find that this solid melts at approximately 3.5 K,  
compared with the experimental melting temperature of 3 K.   
We further find that the commensurate solid phase is 
energetically favored.  At densities below  
commensuration, the system phase 
separates into commensurate solid clusters and a low density vapor.  No 
liquid phase occurs at low temperatures. 
 
\section{Acknowledgments} 
This work was supported in part by the National Aeronautics and Space 
Administration under grant number NAG3-1841.  Some of the calculations 
were performed using the facilities of the Supercomputer Computations 
Research Institute at Florida State University.  We also wish to 
thank L. W. Bruch for useful comments on our work.


\begin{thebibliography}{10}

\bibitem{dash75}
J.~G. Dash, {\em Films on Solid Surfaces} (Academic, New York, 1975).

\bibitem{schick80}
M. Schick,  in {\em Phase Transitions in Surface Films}, edited by J.~G. Dash
  and J. Ruvalds (Plenum, New York, 1980).

\bibitem{bruch97}
L.~W. Bruch, M.~W. Cole, and E. Zaremba, {\em Physical Adsorption: Forces and
  Phenomena} (Oxford, New York, 1997).

\bibitem{bretz77}
M. Bretz, Phys. Rev. Lett. {\bf 38},  501  (1977).

\bibitem{greywall91}
D.~S. Greywall and P.~A. Busch, Phys. Rev. Lett. {\bf 67},  3535  (1991).

\bibitem{greywall93}
D.~S. Greywall, Phys. Rev. B {\bf 47},  309  (1993).

\bibitem{reppy94}
P.~A. Crowell and J.~D. Reppy, Physica B {\bf 187},  278  (1994).

\bibitem{reppy96}
P.~A. Crowell and J.~D. Reppy, Phys. Rev. B {\bf 53},  2701  (1996).

\bibitem{nielsen80}
M. Nielsen, J.~P. McTague, and L. Passell,  in {\em Phase Transitions in
  Surface Films}, edited by J.~G. Dash and J. Ruvalds (Plenum, New York, 1980).

\bibitem{abraham87}
F.~F. Abraham and J.~Q. Broughton, Phys. Rev. Lett. {\bf 59},  64  (1987).

\bibitem{carneiro76}
K. Carneiro {\it et~al.}, Phys. Rev. Lett. {\bf 37},  1695  (1976).

\bibitem{hering76}
S.~V. Hering, S.~W.~V. Sciver, and O.~E. Vilches, J.\ Low Temp.\ Phys. {\bf
  25},  793  (1976).

\bibitem{rabedeau89}
T.~A. Rabedeau, Phys. Rev. B {\bf 39},  9643  (1989).

\bibitem{ecke85}
R.~E. Ecke, Q.-S. Shu, T.~S. Sullivan, and O.~E. Vilches, Phys. Rev. B {\bf
  31},  448  (1985).

\bibitem{bretz73}
M. Bretz, Phys. Rev. Lett. {\bf 31},  1447  (1973).

\bibitem{berker78}
A.~N. Berker, S. Ostlund, and F.~A. Putnam, Phys. Rev. B {\bf 17},  3650
  (1978).

\bibitem{schick77}
M. Schick, J.~S. Walker, and M. Wortis, Phys. Rev. B {\bf 16},  2205  (1977).

\bibitem{tejwani80}
M.~J. Tejwani, O. Ferrerira, and O.~E. Vilches, Phys. Rev. Lett. {\bf 44},  152
   (1980).

\bibitem{crary87}
S.~B. Crary and D.~A. Fahey, Phys. Rev. B {\bf 35},  2102  (1987).

\bibitem{bruch93}
J.~M. Gottlieb and L.~W. Bruch, Phys. Rev. B {\bf 48},  3943  (1993).

\bibitem{mohandas95}
P. Mohandas {\it et~al.}, J.\ Low Temp.\ Phys. {\bf 101},  481  (1995).

\bibitem{pierce99b}
M. Pierce and E. Manousakis, editorially approved for publication in Phys. Rev.
  Lett.

\bibitem{cole80}
W.~E. Carlos and M.~W. Cole, Surf. Sci. {\bf 91},  339  (1980).

\bibitem{whitlock98}
P.~A. Whitlock, G.~V. Chester, and B. Krishnamachari, Phys. Rev. B {\bf 58},
  8704  (1998).

\bibitem{ceprev}
D.~M. Ceperley, Rev. Mod. Phys. {\bf 67},  279  (1995).

\bibitem{pierce99a}
M. Pierce and E. Manousakis, Phys. Rev. B {\bf 59},  3802  (1999)
and Phys. Rev. Lett. {\bf 81} , 156 (1998).

\bibitem{imada84}
M. Takahashi and M. Imada, J. Phys. Soc. Jpn {\bf 53},  3765  (1984).

\bibitem{aziz92}
R.~A. Aziz {\it et~al.}, Mol. Phys. {\bf 77},  321  (1992).

\bibitem{bruch82}
L.~W. Bruch, Surf. Sci. {\bf 125},  194  (1982).

\bibitem{mclachlan64}
A.~D. McLachlan, Mol. Phys. {\bf 7},  381  (1964).

\bibitem{aziz79}
R.~A. Aziz {\it et~al.}, J. Chem. Phys. {\bf 70},  4330  (1979).

\bibitem{whitlock88}
P.~A. Whitlock, G.~V. Chester, and M.~H. Kalos, Phys. Rev. B {\bf 38},  2418
  (1988).

\bibitem{giorgini96}
S. Giorgini, J. Boronat, and J. Casulleras, Phys. Rev. B {\bf 54},  6099
  (1996).

\bibitem{clements93b}
B.~E. Clements, J.~L. Epstein, E. Krotcheck, and M. Saarela, Phys. Rev. B {\bf
  48},  7450  (1993).

\bibitem{gordillo98}
M.~C. Gordillo and D.~M. Ceperley, Phys. Rev. B {\bf 58},  6447  (1998).

\bibitem{bretz78}
S.~E. Polanco and M. Bretz, Phys. Rev. B {\bf 17},  151  (1978).

\bibitem{carneiro81}
K. Carneiro, L. Passell, W. Thomlinson, and H. Taub, Phys. Rev. B {\bf 24},
  1170  (1981).

\bibitem{cole81}
M.~W. Cole and D.~L. Goodstein, Rev. Mod. Phys. {\bf 53},  199  (1981).

\bibitem{derry79}
G.~D. Derry, D. Wesner, W.~E. Carlos, and D.~R. Frankl, Surf. Sci. {\bf 87},
  629  (1979).

\bibitem{carlos80}
W.~E. Carlos and M.~W. Cole, Phys. Rev. B {\bf 21},  3713  (1980).

\bibitem{elgin74}
R.~L. Elgin and L. Goodstein, Phys. Rev. A {\bf 9},  2657  (1974).

\bibitem{ceperley89}
D.~M. Ceperley and E.~L. Pollock, Phys. Rev. B {\bf 39},  2084  (1989).

\bibitem{chen94}
J.~M. Mochel and M.-T. Chen, Physica B {\bf 197},  278  (1994).

\bibitem{boninsegni98}
M. Boninsegni and M.~W. Cole, J.\ Low Temp.\ Phys. {\bf 113},  393  (1998).

\bibitem{cowan98}
J. Ny\'{e}ki, R. Ray, B. Cowan, and J. Saunders, Phys. Rev. Lett. {\bf 81},
  152  (1998).

\bibitem{cheng89}
E. Cheng, G. Ihm, and M.~W. Cole, J.\ Low Temp.\ Phys. {\bf 74},  519  (1989).

\end{thebibliography}
\end{document}